# The Role of Nanotechnology to Combat Major Recent Worldwide Challenges

By:

## G. Ali Mansoori

UNIVERSITY OF ILLINOIS AT CHICAGO
(M/C 063), Chicago, IL 60607-7052, USA
URL: *http://mansoori.people.uic.edu*
EMail: *mansoori@uic.edu*

## Abstract

There are two major challenging problems facing humanity today which are the ongoing Coronavirus disease (COVID-19) pandemic started in late 2019 and the environmental crisis due to fossil fuels combustion causing over increase of greenhouse gases in the atmosphere. While the first problem seems to be a temporary remedy for the second one due to less travels by individuals, actual solution of both problems may require comprehensive scientific, technological, and socioeconomic decisions. In this report we look at the role of nanotechnology and nano-scale materials towards the solution of these two major challenges.

## Recent Problems and Efforts Towards their Solution

### 1. Coronavirus Disease (COVID-19) Pandemic

Since late 2019 one obvious major concern all around the world have been the problems associated with Coronavirus disease (COVID-19) pandemic. This disease is certainly one of the biggest challenges of the 21st century for the whole world. All viruses, including Coronavirus, are nanoscale entities. Coronavirus has a spherical or elliptic shape (Figure 1), and it is said to also assume pleomorphic (irregular and variant) form at various conditions. It has a diameter of about 60–140 nm [*www.ncbi.nlm.nih.gov › books › NBK55477*] which is close to the range of nano-scale (1-100 nm) sizes [Mansoori, 2005].

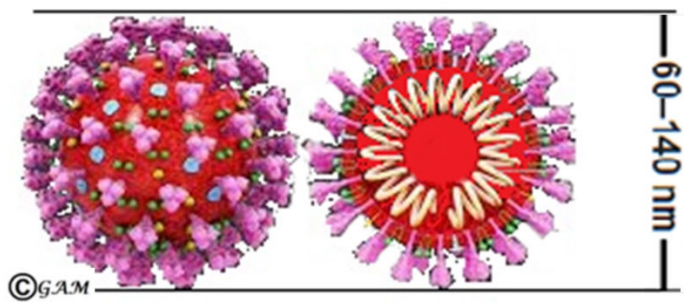

**Figure 1.** Coronavirus is a nanoscale entity

It is not surprising that there has been a great deal of interest in finding applications of the principles of nanotechnology combined with biology to combat the virus which causes this disease. In the literature related to developing mRNA COVID-19 vaccines, it is stated that lipid nanoparticles, which had been in the works for decades, is used as delivery vehicles, accelerating the development of these vaccines to save humans' lives [Cross 2021]. It is known that, these medical interventions use nanotechnology to mimic nature's own method of slipping past the immune system to deliver treatment to target cells [Kulkarni, et al., 2020; Shin, et al., 2020]. It is recommended that this technology for delivering vaccines into the human body should be considered by researchers, if the gene editing system for genetic disease treatment using CRISPR (Clustered Regularly Interspaced Short Palindromic Repeats) tool [Kato-Inui, et al., 2018] is to have any success in the future. While such scientific efforts seem quite useful in the design of vaccines, there is also a need for reliability studies of the vaccine effectiveness, similar to what has been done for nano-drug delivery [Ebrahimi and Mansoori, 2014; Ebrahimi, et al, 2016], in regard to its duration of produced immunity and the possible requirement for its repetition or booster shots.

In addition to efforts to develop vaccines based on nanotechnology, there have been other recent nanotechnology-based efforts to make people more immune from COVID-19, the current most important public health problem. In addition to recommendation for social distancing we are encouraged wearing facial and possibly hands protection (Figure 2).

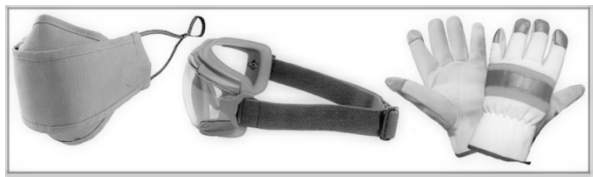

**Figure 2.** Facial and hands protection covers

Such requirement against COVID-19 and other surfaces and airborne pathogens, brought about the design and production of nanotechnology-based facial and hands protection. Nanofibers and nanoparticles have been considered for the design of some face masks to improve their antiviral performance. With the discovery of variety of nanoparticles (carbon nanotubes, diamondoids, fullerenes, graphene, gold nanoparticles, quantum dots, silver nanoparticles, titanium dioxide nanostructures, etc.), there lies a vast field of unsolved medical diagnoses to be reassessed. Before application of nanoparticles in the design of face masks and other protective devices, in addition to their role in killing the bacteria and viruses, it is important to know what happens to a particle once it is free in the body [Hartung and Mansoori, 2013].

Considering the popularity of graphene, a novel nanomaterial, which has found interesting applications specially in opto-electronics area [Kovalev, et al., 2021; Liu et al., 2021], some groups have also considered graphene as a novel nanomaterial which possesses antiviral and antibacterial properties. Then, there were face masks designed, manufactured and sold containing graphene with protection against COVID-19 claims and used by adults and children in certain schools and daycares. However, on April 2, 2021, Health Canada, the department of the Government of Canada responsible for national health policy, issued an advisory mentioning "Face masks that contain graphene may pose health risks". But on July 13, 2021 issued another advisory update mentioning "Graphene face masks by Shandong Shengquan New Materials Co. Ltd. can resume sale in Canada; Health Canada found no health risks of concern with these products" [*Face masks that contain graphene may pose health risks - Recalls and safety alerts (healthycanadians.gc.ca)*]. There are still this question "whether graphene-coated face masks a COVID-19 miracle – or another health risk?" by health professionals [*theconversation.com/are-graphene-coated-face-masks-a-covid-19-miracle-or-another-health-risk-159422*]. That is because there is a potential to inhale graphene particles present in the mask, which may pose health risks. Carbon atoms at the edges of a graphene sheet (Figure 3) have special chemical reactivity if they are not properly reacted. Graphene has the highest ratio of edge atoms of any allotrope, including its own allotropes which include carbon nanotube, nano-diamonds, and Fullerene.

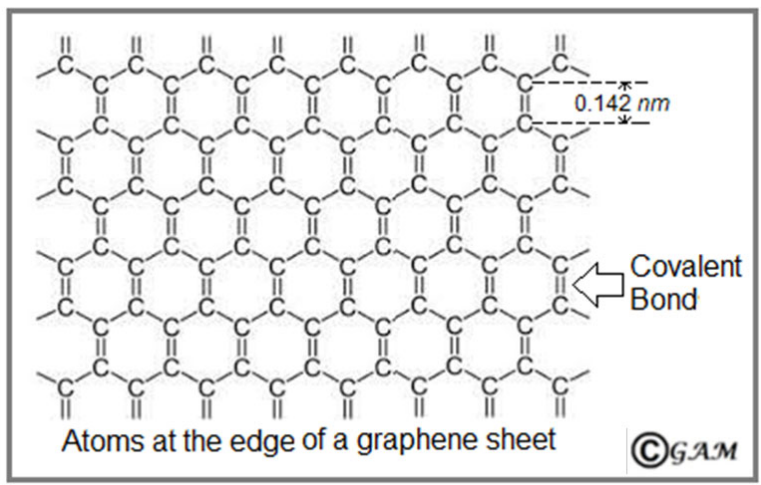

**Figure 3.** Graphene has the highest ratio of edge atoms of any allotrope

Previous studies have shown pristine or unmodified carbon nanotubes can cause pulmonary fibrosis/lung scarring and lung inflammation in animals and cellular models after inhalation exposure in a similar way as asbestos [Gracin, et al., 2009]. However, studies indicate graphene could one day be used as an electronic biosensor to make quick, reliable tests for viruses like SARS-CoV-2. In a recent review [Prattis, et al., 2021] investigators looked into the latest research to find the most exciting potential applications of graphene in point-of-care tests. This includes diagnostic tests for the virus responsible for COVID-19, but also detecting other viruses, bacteria and even cancerous tumors.

      It should be mentioned that various investigations have indicated that $TiO_2$ nanostructures [Khataee and Mansoori, 2011] and silver nanoparticles [Vahabi, et al., 2011; Mohammadinejad, et al., 2013], among other nanosystems are also effective to combat the bacteria and viruses. Consideration of these and other nanostructures in the design of protective human shields require carefully analyzed in terms of efficacy and possible long-term effects on the wearers' skin and lungs as well as on the environment [Palmieri, et al., 2021].

## 2. Environmental Crisis due to Fossil Fuels Combustion

While all the countries around the world are trying to combat the COVID-19 pandemic, the earth environment is being damaged due to the excessive amount of fossil fuels combustion and oversupply of greenhouse gases in the atmosphere [Mohammed and Mansoori, 2017]. This has made many industrially advanced societies to consider abandoning fossil fuels for combustion and utilize the renewable energies. The 2015 United Nations Climate Change Conference (COP21 Paris Agreement (Figure 4) provided fundamental guidelines to achieve energy sufficiency, clean environment, and sustainable sources for countries, states, or communities.

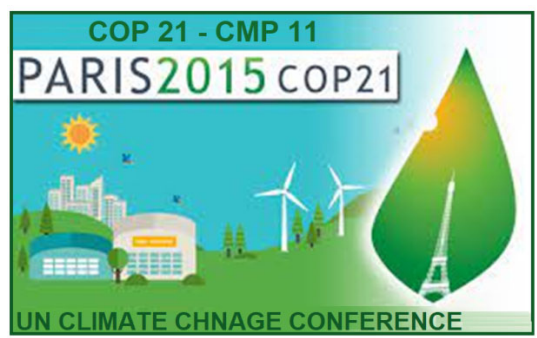

**Figure 4.** The 11[th] Session of the Conference of the Parties serving as the meeting of the Parties to the Kyoto Protocol (CMP 11) took place from Monday, 30 Nov. to Friday, 11 Dec. 2015 in Paris le Bourget, France.

We published a comprehensive book covering the subject of energy sources, utilization, legislation, and sustainability a year after the COP21 Paris Agreement, which represents a roadmap to follow its requirements [Mansoori et al., 2016]. In the book there are discussions on the role of nanotechnology and nanoparticles in improving the sustainability of the energy industry. Several industrially advanced countries have plans to abandon the use of fossil fuels combustion and utilize renewable energies. While the technology for small-scale renewable energies utilization is already available, for its large-scale utilization, storage and transport of appropriate fuels are required.

In a recent publication [Mansoori, et al., 2021] we have studied the potential of various non-fossil/alternative fuels for large-scale renewable energy sources utilization. Our studies indicate that production of biofuels using renewable energies will allow biofuels direct use in the existing internal combustion engines. Then, there may be no need for major electrification of the transportation industry which will require huge amount of expensive and rare materials, such as lithium, and silver to have high-capacity batteries [Mansoori, et al., 2021]. However, for the agricultural activities based on which biofuels are produced there is a need for ammonia, an important agro-chemical since it is the major feed for agricultural fertilizers. It is interesting to note that ammonia can be also produced using renewable energies as it is demonstrated in Figure 5. However, there are nanotechnology approaches being studied for improvement in its production conditions from renewable energies [Li, et al., 2019; Sun, et al., 2020]. Also, in production of ammonia, an important agro-chemical, from renewable sources of energy, hydrogen is also needed to be produced using renewable energies (Figure 5).

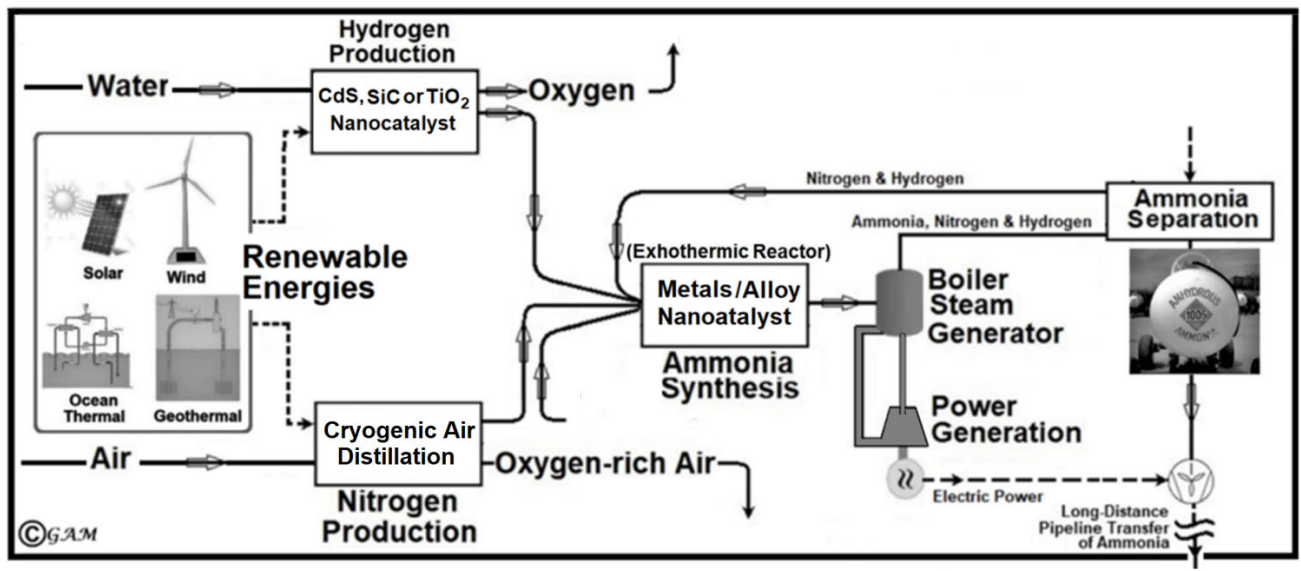

**Figure 5.** Hydrogen production through dissociation of water, nitrogen purification through cryogenic distillation of air and ammonia (an important agro-chemical) production, storage and transport, all through the use of renewable energies

There are several efforts underway to apply certain nanomaterials for improvement of hydrogen production from renewable energies, as well as its storage and transport in case of using it as a fuel in the future [Li, 2013; Chen, et al., 2021; Sen, et al., 2021]. The technology for storage and transport of ammonia is already available worldwide.

The book on "Nano-enabled Agrochemicals in Agriculture" [M. Ghorbanpour and MA Shahid (Ed's), 2021], to which this essay is the introductory chapter, contains 29 papers covering various aspects of applications of nanotechnology towards improvement of agricultural activity, including biofuels production.